# Factors Determining Nestedness in Complex Networks

Samuel Jonhson[1], Virginia Domínguez-García[2], Miguel A. Muñoz[2]*

1 Department of Mathematics, Imperial College, London, United Kingdom, 2 Departamento de Electromagnetismo y Física de la Materia e Instituto Carlos I de Física Teórica y Computacional, Universidad de Granada, Granada, Spain

**Abstract**

Understanding the causes and effects of network structural features is a key task in deciphering complex systems. In this context, the property of *network nestedness* has aroused a fair amount of interest as regards ecological networks. Indeed, Bastolla *et al.* introduced a simple measure of network nestedness which opened the door to analytical understanding, allowing them to conclude that biodiversity is strongly enhanced in highly nested mutualistic networks. Here, we suggest a slightly refined version of such a measure of nestedness and study how it is influenced by the most basic structural properties of networks, such as degree distribution and degree-degree correlations (i.e. assortativity). We find that most of the empirically found nestedness stems from heterogeneity in the degree distribution. Once such an influence has been discounted – as a second factor – we find that nestedness is strongly correlated with disassortativity and hence – as random networks have been recently found to be naturally disassortative – they also tend to be naturally nested just as the result of chance.

**Funding:** This work was supported by Junta de Andalucia projects FQM-01505 and P09-FQM4682, and by Spanish MEC-FEDER project FIS2009-08451. S.J. is grateful for financial support from the European Commision under the Marie Curie Intra-European Fellowship Programme PIEF-GA-2010-276454. The funders had no role in study design, data collection and analysis, decision to publish, or preparation of the manuscript.

**Competing Interests:** The authors have declared that no competing interests exist.

* E-mail: mamunoz@onsager.ugr.es

## Introduction

Networks have become a paradigm for understanding systems of interacting objects, providing us with a unifying framework for the study of diverse phenomena and fields, from molecular biology to social sciences [1]. Most real networks are not assembled randomly but present a number of non-trivial structural traits such as the small-world property, scale freeness, hierarchical organization, etc [2,3]. Network topological features are essential to determine properties of complex systems such as their robustness, resilience to attacks, dynamical behavior, spreading of information, etc. [3–5]. A paradigmatic case is that of ecosystems, in which species can be visualized as nodes of a network and their mutual interactions (predation, mutualism, etc) encoded in the edges or links. In this context, the solution to May's famous paradox [6] – the fact that large ecosystems seem to be especially stable, while random matrix theory predicts the contrary – is still not fully clear, but it is widely suspected that there are structural (non random) features of ecological networks at the basis of enhanced stability, which as yet elude us (see [7] for a recent challenge to this idea).

One such feature of ecological networks, which has been studied for some time by ecologists, is called *nestedness* [8]. Loosely speaking, a bipartite network [3] – say, for argument's sake, of species and islands, linked whenever the former inhabits the latter – is said to be nested if the species that exist on a few islands tend always to be found also on those islands inhabited by many different species. This can be most easily seen by graphically representing a matrix such that species are columns and islands are rows, with elements equal to one whenever two nodes are linked and zero if not. If, after ordering all nodes by degree (number of neighbours), most of them can be quite neatly packed into one corner, the network is considered highly nested [8,9]. This is illustrated in Fig. 1 where we plot different connectivity matrices with different levels of maximal "compactability" and, thus, with different levels of nestedness.

Nestedness is usually measured with purposely-designed software. The most popular *nestedness calculator* is the "temperature of Atmar and Patterson (used to extract a temperature from the matrices in Fig. 1) [8]. It estimates a curve of equal density of ones and zeros, calculates how many ones and zeros are on the "wrong" side and by how much, and returns a number between 0 and 100 called "temperature" by analogy with some system such as a subliming solid. A low temperature indicates high nestedness. It is important to caution that nestedness indices should not be used as black-boxes, as this can lead to false conclusions [10,11]. The main drawback of these calculators is that they are defined by complicated algorithms, hindering further analytical developments. Even if initially introduced for bipartite networks, the concept of nestedness can be readily generalized for generic networks.

In a seminal work, Bascompte and collaborators [12] showed that real *mutualistic networks* (i.e. bipartite networks of symbiotic interactions), such as the bipartite network of plants and the insects that pollinate them, are significantly nested. They also defined a measure to quantify the average number of shared partners in these mutualistic networks, and called it "nestedness" because of its close relation with the concept described above. They go on to show evidence of how the so-defined nestedness of empirical mutualistic networks is correlated with the biodiversity of the corresponding ecosystems [13]: the global species competition is significantly reduced by developing a nested network architecture and this entails a larger biodiversity. The principle behind this is simple. Say nodes A and B are in competition with each other. An increase in A will be to B's detriment and vice-versa; but if both A and B engage in a symbiotic relationship with node C, then A's thriving will stimulate C, which in turn will be helpful to B. Thus, the effective competition between A and B is reduced, and the whole system becomes more stable and capable of sustaining more nodes and more individuals. The beneficial effect that "competing" nodes (i.e. those in the same side of a bipartite network) can gain from sharing "friendly" partners (nodes in the other side) is not confined to ecosystems. It is expected also to play a role, for instance, in financial networks or other economic systems [14]. To what extent the measure introduced by Bascompte et al. is related to the traditional concept of nestedness has not, to the best of our knowledge, been rigorously explored. Irrespectively of this relation, however, the insight that mutual neighbours can reduce effective competition in a variety of settings is clearly interesting in its own right, and it is for this reason that we analyse this feature here. On a different front, Staniczenko et al. [15] have made some promising analytical progress regarding the traditional concept of nestedness.

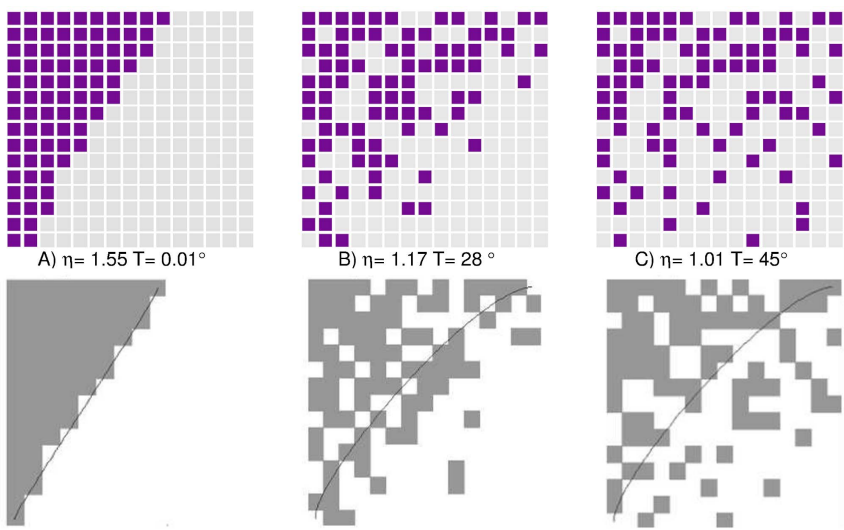

**Figure 1. Measures of nestedness in networks.** The figure shows three different connectivity matrices with different levels of nestedness as measured by (i) our new nestedness index [Eq. (6)] and (ii) the standard nestedness "temperature' calculator". As can be readily seen, the most packed matrix corresponds to a very low temperature and to a high nestedness index ($\eta > 1$) and, reciprocally, the least packed one exhibits a high temperature and an index close to its expected value for a random network ($\eta \simeq 1$).
doi:10.1371/journal.pone.0074025.g001

Here, we take up this idea of shared neighbours (though characterized, owing to reasons we shall explain in the Methods, with a slightly different measure) and study analytically and computationally how it is influenced by the most relevant topological properties, such as the degree distribution and degree-degree correlations. Our aim is to understand to what extent nestedness is a property inherited from imposing a given degree distribution or a certain type of degree-degree correlations.

## Methods: Analytical Quantification of Nestedness

Consider an arbitrary network with $N$ nodes defined by the adjacency matrix $\hat{a}$: the element $\hat{a}_{ij}$ is equal to the number of links (or edges) from node $j$ to node $i$ (typically considered to be either 1 or 0 though extensions to weighted networks have also been considered in the literature [15]). If $\hat{a}$ is symmetric, then the network is undirected and each node $i$ can be characterized by a degree $k_i = \sum_j \hat{a}_{ij}$. If it is directed, $i$ has both an *in* degree, $k_i^{\text{in}} = \sum_j \hat{a}_{ij}$, and an *out* degree, $k_i^{\text{out}} = \sum_j \hat{a}_{ji}$; we shall focus here on undirected networks, although most of the results could be easily extended to directed ones.

Bastolla *et al.* [13] have shown that the effective competition between two species can be reduced if they have common neighbours with which they are in symbiosis. Therefore, in mutualistic networks it is beneficial for the species at two nodes $i$ and $j$ if the number of shared symbiotic partners, $\hat{n}_{ij} = \sum_l \hat{a}_{il}\hat{a}_{lj} = (\hat{a}^2)_{ij}$, is as large as possible. Going on this, and assuming the network is undirected, the authors propose to use the following measure:

$$\eta_B = \frac{\sum_{i<j} \hat{n}_{ij}}{\sum_{i<j} \min(k_i, k_j)}, \qquad (1)$$

which they call *nestedness* because it would seem to be highly correlated with the measures returned by nestedness software. Note that, although the authors consider only bipartite graphs, such a feature is not imposed in the above definition.

Here, we take up the idea of the importance of having an analytical expression for the nestedness but, for several reasons, we use a definition slightly different from the one in [13]. Actually, $\eta_B$ suffers from a serious shortcoming; if one commutes the sums in the numerator of Eq. (1), it is found that the result only depends on the heterogeneity of the degree distribution: $\sum_{ij} \hat{n}_{ij} = \sum_l \sum_i \hat{a}_{il} \sum_j \hat{a}_{lj} = N\langle k^2 \rangle$ (in an undirected network, $\sum_{i<j} = \frac{1}{2}\sum_{ij}$; we shall always sum over all $i$ and $j$, since it is easier to generalize to directed networks and often avoids writing factors 2). Therefore, this index essentially provides a measurement of network heterogeneity. Also, although the maximum value $\hat{n}_{ij}$ can take is $\min(k_i, k_j)$, this is not necessarily the best normalization factor, since (as we show explicitly in the next Section) the randomly expected number of paths of length 2 connecting nodes $i$ and $j$ depends on both $k_i$ and $k_j$. Furthermore, it can sometimes be convenient to have a local measure of nestedness (i.e. nestedness of any given node) which cannot be inferred from the expresion above. For all these reasons, we propose to use

$$\tilde{\eta}_{ij} \equiv \frac{\hat{n}_{ij}}{k_i k_j} = \frac{(\hat{a}^2)_{ij}}{k_i k_j}, \quad (2)$$

which is defined for every pair of nodes $(i,j)$. This allows for the consideration of a nestedness per node, $\tilde{\eta}_i = N^{-1} \sum_j \tilde{\eta}_{ij}$, or of the global measure

$$\tilde{\eta} = \frac{1}{N^2} \sum_{ij} \tilde{\eta}_{ij} \quad (3)$$

which is very similar in spirit to the measure introduced by Bastolla et al. in [13] but, as argued above, has a number of additional advantages. This new index can be easily applied to bipartite networks, as shown in Appendix S1.

Having an analytical definition of nestedness, it becomes feasible to scrutinize how it is influenced by the most basic structural features, such as the degree distribution and degree-degree correlations. The standard procedure to determine how significantly nested a given network is, is to generate randomizations of it (while keeping fixed some properties such as the total number of nodes, links, or degree distribution) and compare the nestedness of the initial network with the ensemble-averaged one. The set of features kept fixed in randomizations determine the *null-model* used as reference.

### Effects of the Degree Distribution: Configuration Model

Many networks have quite broad degree distributions $P(k)$; most notably the fairly ubiquitous scale-free networks, $P(k) \sim k^{-\gamma}$ [2]. Since heterogeneity tends to have an important influence on any network measure, it is important to analytically quantify the influence of degree-distributions on nestedness. For any particular degree sequence, the most natural choice is to use the *configuration model* [3,16] – defined as the ensemble of random networks wired according to the constraints that a given degree sequence $(k_1,...,k_N)$ is respected – as a *null model*. In such an ensemble, the averaged value of any element of the adjacency matrix is

$$\overline{\hat{a}_{ij}} \equiv \hat{y}^c = \frac{k_i k_j}{\langle k \rangle N}. \quad (4)$$

We use an overline, $\overline{(\cdot)}$, to represent ensemble averages and angles, $\langle \cdot \rangle$, for averages over nodes of a given network.

### Nestedness in the Configuration Model

Plugging Eq. (4) into Eq. (2), we obtain the expected value of $\tilde{\eta}$ in the configuration ensemble, which is our basic null model

$$\overline{\tilde{\eta}_{ij}} = \frac{\langle k^2 \rangle}{\langle k \rangle^2 N} \equiv \tilde{\eta}_{conf}. \quad (5)$$

It is important to underline that $\overline{\tilde{\eta}_{i,j}}$ is independent of $i$ and $j$; hence, it coincides with the expected value for the global measure, $\overline{\tilde{\eta}} = \overline{\tilde{\eta}_{i,j}}$ (which justifies the normalization chosen in Eq. (2)). Also, it is noteworthy that for degree distributions with finite first and second moments, $\tilde{\eta}_{conf}$ goes to zero as the large-N limit is approached.

It is obvious from Eq. (5) that degree heterogeneity has an important effect on $\tilde{\eta}$; for instance, scale-free networks (with a large degree variance) are much more nested than homogeneous ones. Therefore, if we are to capture aspects of network structure other than those directly induced by the degree distribution it will be useful to consider the nestedness index normalized to this expected value,

$$\eta \equiv \frac{\tilde{\eta}}{\tilde{\eta}_{conf}} = \frac{\langle k \rangle^2}{\langle k^2 \rangle N} \sum_{ij} \frac{(\hat{a}^2)_{ij}}{k_i k_j}. \quad (6)$$

Although $\eta$ is unbounded, it has the advantage that it is equal to unity for any uncorrelated random network, independently of its degree heterogeneity, thereby making it possible to detect additional non-trivial structure in a given empirical network.

### Degree-degree Correlations in the Configuration Model

In the configuration ensemble, the expected value of the mean degree of the nearest neighbours (nn) of a given node is $\overline{k_{nn,i}} = k_i^{-1} \sum_j \hat{y}^c k_j = \langle k^2 \rangle / \langle k \rangle$, which is independent of $k_i$. Still, specific finite-size networks constructed with the configuration model can deviate from the ensemble average results (which hold exactly only in the $N \to \infty$ limit). Real networks are finite, and they often display degree-degree correlations, which result in $\overline{k_{nn,i}} = \overline{k_{nn}}(k_i)$. If $\overline{k_{nn}}(k)$ increases (decreases) with $k$, the network is said to be assortative (disassortative), i.e. nodes with large degree tend to be connected with other nodes of large (small) degree.

The measure usually employed of this phenomenon is Pearson's coefficient applied to the edges [3,4,17]: $r = ([k_l k'_l] - [k_l]^2)/([k_l^2] - [k_l]^2)$, where $k_l$ and $k'_l$ are the degrees of each of the two nodes belonging to edge $l$, and $[\cdot] \equiv (\langle k \rangle N)^{-1} \sum_l (\cdot)$ is an average over edges. Writing $\sum_l (\cdot) = \sum_{ij} \hat{a}_{ij}(\cdot)$, $r$ can be expressed as [17]

$$r = \frac{\langle k \rangle \langle k^2 \overline{k_{nn}}(k) \rangle - \langle k^2 \rangle^2}{\langle k \rangle \langle k^3 \rangle - \langle k^2 \rangle^2}. \quad (7)$$

In the infinite network-size limit we expect $r = 0$ in the configuration model (null model) as there are no built in correlations. Even if the index $r$ is widely used to measure network correlations, some drawbacks of it have been put forward [18,19].

## Results

### Emergence of Effective Correlations in Finite-size Networks

We have computationally constructed finite random networks with different degree distributions; in particular, Poissonian, Gaussian, and scale-free distributions, assembled using the configuration model as explained above (for the scale-free case see Ref. [20]) and measured their Pearson's correlation coefficient. Results are illustrated in Fig. 2; the probability of obtaining negative (disassortative) values of $r$ is larger than the one for positive (assortative) values (observe the shift between $r = 0$ and the curve averaged value). This means that the null-model expectation value of $r$ is negative! i.e. finite random networks are more likely to be disassortative than assortative. This result is highly counterintuitive because the ensemble is constructed without assuming any type of correlations and is, clearly, a finite-size effect. Indeed, for larger network sizes the averaged value of $r$ converges to 0 as we have analytically proved and computationally verified. For instance, for scale-free networks, $r$ can be easily shown to converge to 0 as $r \propto N^{-1/3}$ in the large-$N$ limit (see Appendix S2 and Fig. 2B). A well-known effect leading to effective disassortativity, is that simple algorithms, which are supposed to generate uncorrelated networks, can instead lead to degree-degree anti-correlations when the desired degree distribution has a heavy tail and no more than one link is allowed between any two vertices (as hubs are not as connected among themselves as they should be without such a constraint) [21,22]. Also, our observation is in agreement with the recent claim that, owing to entropic effects, real scale-free networks are typically disassortative: simply, there are many more ways to wire networks with disassortative correlations than with assortative ones [23].

### Effective Correlations Imply Nestedness in Finite Networks

A straightforward consequence of the natural tendency of finite networks to be disassortative is that they thereby also become naturally nested. Indeed, the nestedness index $\eta$ was defined assuming there were no built-in correlations, but if degree-degree correlations effectively emerge in finite-size random networks, then deviations from the neutral value $\eta = 1$ are to be expected. Indeed, in Fig. 2C we have considered networks constructed with the configuration model, employing the same probability distributions (Gaussian, Poissonian and scale free) as above. For each so-constructed random network we compute both $r$ and $\eta$ and plot the average of the second as a function of the first (technical details on how to sample networks with extreme values of $r$ – using the Wang-Landau algorithm [24] – are given in Appendix S3). The resulting three curves exhibit a neat (almost linear) dependence of the expected value of $\eta$ on $r$: disassortative networks are nested while assortative ones are anti-nested. As disassortative ones are more likely to appear, a certain degree of nestedness is to be expected in finite random networks. Observe that for truly uncorrelated random networks, i.e. with $r = 0$, the expectation value of $\eta$ is 1.

Finally, in Appendix S4, we provide an analytical connection between disassortativity and nestedness in random networks with explicitly built-in degree-degree correlations. Also in this case a clear relation between nestedness and disassortativity emerges (as shown in the figure of Appendix S4) for scale-free networks.

### Degree Correlations in Real vs Randomized Networks

We have considered 60 different empirical networks, both bipartite and unimodal, from the literature. The set includes foodwebs, metabolic, neuronal, ecological, social, and technological networks (see Appendix S5). We have performed randomizations preserving the corresponding degree sequences (configuration ensemble) and avoiding multiple links between any pair of nodes. Results for a subset of 16 networks are illustrated in Figure 3, which shows the distribution of r-values (see figure caption) compared with the actual value of $r$.

The actual value of $r$ in empirical networks coincides with the ensemble average within an error of the order of 1, 2, or 3 standard deviations in about two thirds of the cases (53%, 67%, and 76% respectively). Similarly, the corresponding p-values are larger than the significance threshold (0.05) in 60% of the cases. Particularizing for bipartite networks, the z-scores rise to: 60%, 76%, and 89%, respectively, and the significant P-values go up to 68% (data are collected in Appendix S5).

Therefore, roughly speaking, the null model – in which networks are randomly wired according to a specified degree sequence – explains well the correlations of about two-thirds (or more) of the networks we have analysed and, more remarkably, it explains even better the correlations of bipartite networks. Thus, once it has been realized that random networks have a slight natural tendency to be disassortative, in many cases, there does not seem to be a clear generic statistical tendency for real networks to be more correlated (either assortatively or disassortatively) than expected in the null model. For instance in almost all foodwebs we have analyzed the empirical value of $r$ is well explained by randomizations, while in some other social and biological networks there are some residual positive correlations (assortativity).

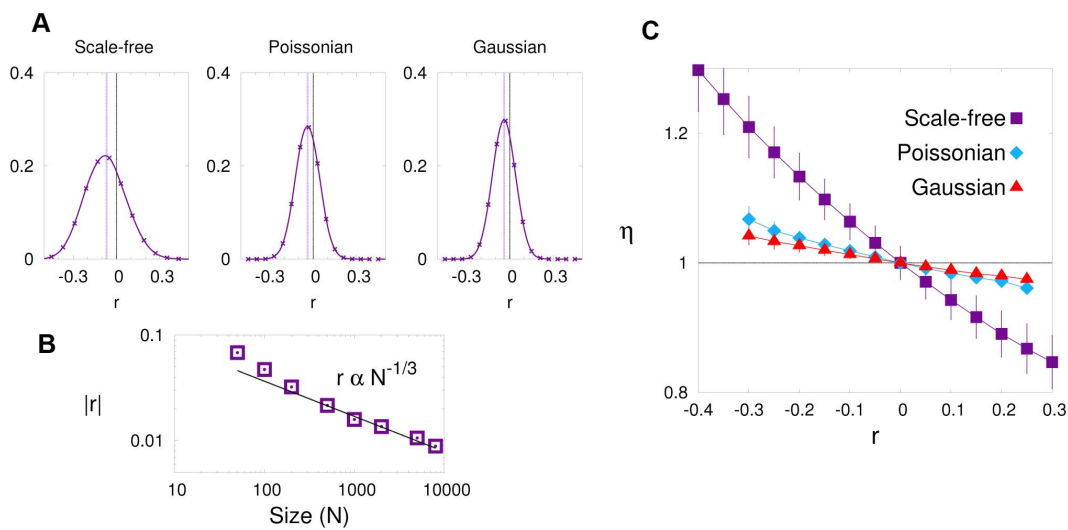

**Figure 2. Correlation coefficient and nestedness in random networks.** (Panel A): Correlation coefficient, $r$, and nestedness $\eta$ for $10^6$ networks generated independently using the configuration model with $N = 50$ nodes and $<k> = 5$ and (from left to right) scale-free (with exponent $\gamma = 2.25$), Poissonian, and Gaussian ($\sigma^2 = 10$) degree distributions. (Panel B): Pearson's correlation coefficient as a function of network size for scale free networks with $\gamma = 2.25$. (Panel C): Averaged nestedness (with error bars corresponding to one standard deviation) as a function of Pearson's correlation index $r$ in random (scale-free, Poissonian, and Gaussian) networks (as in the left panel). These curves are obtained employing the Wang-Landau algorithm as described in Appendix S3. All three curves show a positive (almost linear) correlation between disassortativity and nestedness: more disassortative networks are more nested. By restricting the corresponding configuration ensembles to their corresponding subsets in which $r$ is kept fixed it is possible to define a more constraint null model as discussed in the main text.
doi:10.1371/journal.pone.0074025.g002

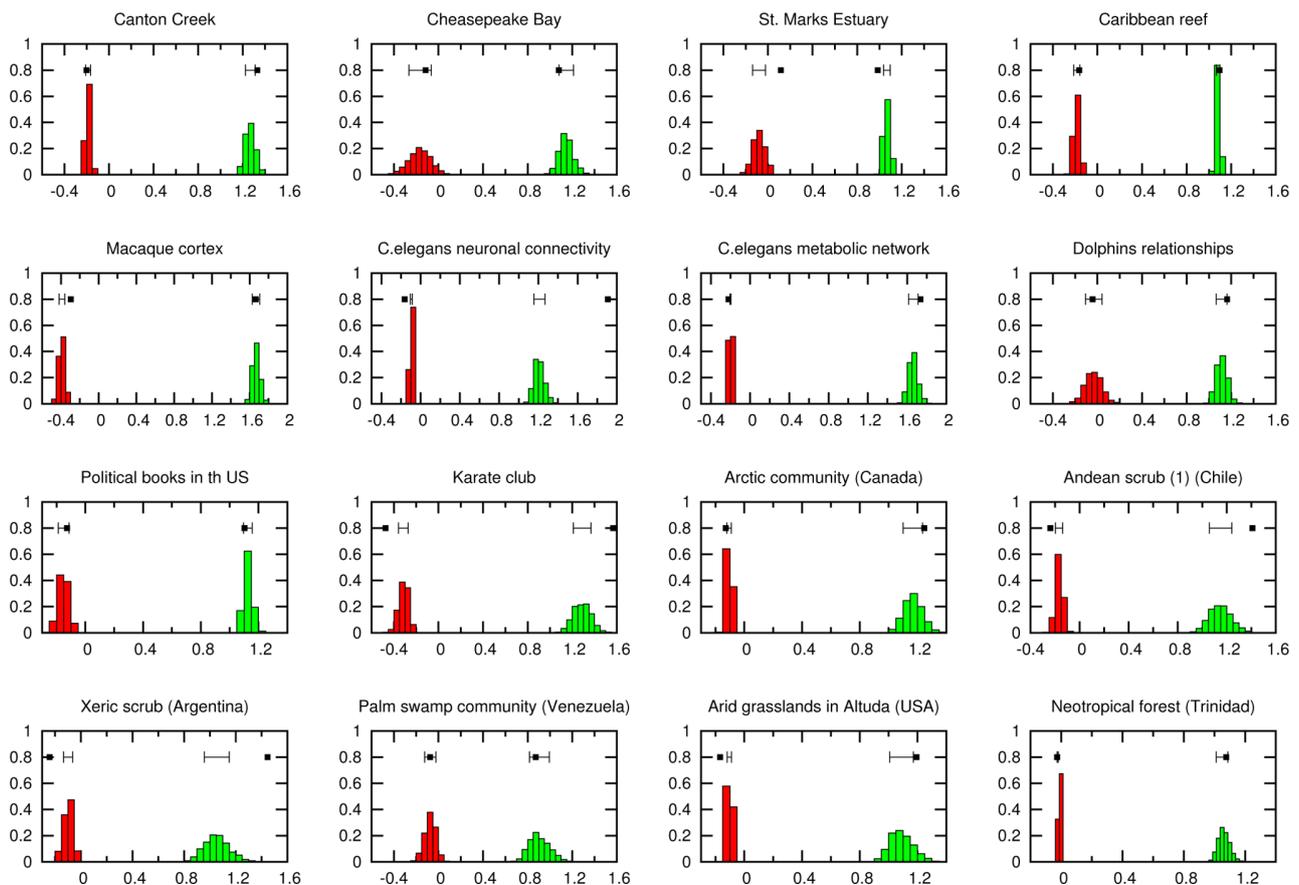

**Figure 3. Correlation coefficient and nestedness in degree-preserving randomiaztions.** Probability distribution of Pearson's coefficient $r$ and of the nestedness coefficient, $\eta$, as measured in degree-preserving randomizations of a subset of 16 (out of a total of 60) real empirical networks (as described and referenced in Appendix S5). The actual empirical values in the real network are marked with a black box and compared (also in black) with a segment centered at the mean value of the random ensemble (configuration model) with width equal to one standard deviation. In most cases but not all, the empirical values lie in or near the corresponding interval, suggesting that typically empirical networks are not significantly more assortative/nested than randomly expected.
doi:10.1371/journal.pone.0074025.g003

### Nestedness in Real vs Randomized Networks

We have conducted a similar analysis for the nestedness index $\eta$ and compare its value in real networks with the expected value in randomizations (see Fig. 3). In this case, the actual value of $\eta$ in empirical networks coincides with the ensemble average with an error of the order of 1, 2, or 3 standard deviations also in about two thirds of the cases (43%, 73%, and 83% respectively). As for the p-value, it is above threshold in 63% of the cases (which goes up to 76% for bipartite networks). Thus, in most of the analysed examples, empirically observed values of nestedness are in agreement with null-model expectations once the degree-distribution has been taken into consideration (data shown in Appendix S5).

### Nestedness vs Degree Correlations in Empirical Networks

As said above, both Fig. 2C and Fig. 3 reveal a global tendency: exceedingly disassortative empirical networks tend to be nested while assortative ones are anti-nested. To further explore this relation, Fig. 4 shows a plot of nestedness against assortativity for the selection of empirical networks listed in Appendix S5. Although these networks are highly disparate as regards size, density, degree distribution, etc., it is apparent that the main contribution to $\eta$ comes indeed from degree-degree correlations. *The observation of such a strong generic correlation between the nestedness and disassortativity constitutes one of the main findings of this paper.*

### A more Refined Null Model

A unique criterion for choosing a proper null model does not exist [25]. For instance, it is possible to go beyond the null model studied so far by preserving not just the degree sequence but also empirical correlations. Indeed, from the full set of networks generated with the configuration model for a given degree sequence, one could consider the subset of networks with a fixed value of $r$, as done in Fig. 2C (and as explained in Appendix S3). In particular, one could take the sub-ensemble with the same $r$ as empirically observed. This constitutes a more refined null model in which the number of nodes, degree sequence, and degree-degree correlations are preserved. This more refined null model reproduces slightly better than the configuration model the empirical values of nestedness; for instance, allowing for three standard deviations bipartite networks are explained in a 100% of the cases (details can be found in Appendix S3). Thus, the null model preserving degree-degree correlations explains quite well the observed levels of nestedness.

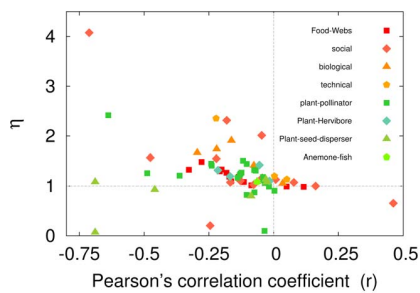

**Figure 4. Nestedness against assortativity (as measured by Pearson's correlation coefficient) for data on a variety of networks.** Warm-coloured items correspond to unimodal networks and green ones to bipartite networks of different kinds (see Appendix S5).
doi:10.1371/journal.pone.0074025.g004

## Discussion and Conclusions

Theoretical studies suggest that a nested structure minimizes competition and increases the number of coexisting species [13], and also it makes the community more robust to random extinctions [26] and habitat loss [27]. In order to make progress, systematic analyses of nestedness and nestedness indices are necessary.

The first contribution of this work is that a new analytical nestedness index has been introduced. It is a variant of the one introduced in Ref. [13], allowing for analytical developments, which are not feasible with standard computational estimators (or calculators) of nestedness. Besides that, the new index exhibits a number of additional advantages: (i) it allows us to identify the amount of nestedness associated with each single node in a network, making it possible to define a "local nestedness"; (ii) the new index is properly normalized and provides an output equal to unity in uncorrelated random networks, allowing us in this way to discriminate contributions to nestedness beyond network heterogeneity.

Having removed the direct effects of the degree distribution – which has a dominant contribution to other measures of nestedness – it is possible to move one step forward and ask how degree-degree correlations (as quantified by Pearson's coefficient) influence nestedness measurements. Curiously enough, there are more disassortative (negatively degree-degree correlated) networks than assortative ones even among randomly assembled networks. Different reasons for this have already being pointed out in the literature [21–23] and we have confirmed that indeed this is the case for finite networks built with the configuration model.

Therefore, the neutral expectation for finite random networks is to have some non-vanishing level of disassortativity ($r<0$). Analogously, as we have first reported here, there is a very similar tendency for finite random networks to be naturally nested. There is a clean-cut correspondence between nestedness and disassortativity: disassortative networks are typically nested and nested networks are typically disassortative. This is true for finite-size computational random models, analytically studied correlated networks of any size (Appendix S4), as well as in real empirical networks (as vividly illustrated in Figure 2C and Fig. 4).

Analyses of 60 empirical networks (both bipartite and non-bipartite) taken from the literature reveal that in many cases the measured nestedness is in good correspondence with that of the degree-preserving null model. In particular, almost 90% of the studied bipartite networks are well described by the null model and this figure rises up to 100% when a more refined null model is considered. Finally, recent results by Allesina's group [15] suggest that one should consider weighted networks to properly study nestedness; we leave an extension of our analyses along this line for a future work.

In conclusion, degree heterogeneity together with the finite size of real networks suffice to justify most of the empirically observed levels of nestedness in ecological bipartite network.

## Supporting Information

**Appendix S1** In this appendix we show how to generalize the new nestedness index to bipartite networks.
(PDF)

**Appendix S2** This appendix explains how the Pearson's correlation coefficient scales with size in finite scale-free networks.
(PDF)

**Appendix S3** This appendix illustrates how to sample networks with a given value of the Pearson's correlation coefficient.
(PDF)

**Appendix S4** In this appendix we analytically compute degree-degree correlations in heterogeneous networks.
(PDF)

**Appendix S5** This appendix contains tables with the network data used in the manuscript.
(PDF)

## Acknowledgments

We are especially grateful to J.A. Dunne for providing us with food-web data, and to A. Arenas and M. Newman for making the other network data available online. Many thanks also to A. Pascual-García and P. Moretti for fruitful conversations.

## Author Contributions

Conceived and designed the experiments: SJ VDG MAM. Performed the experiments: SJ VDG MAM. Analyzed the data: SJ VDG MAM. Contributed reagents/materials/analysis tools: SJ VDG MAM. Wrote the paper: SJ VDG MAM.

## References

1. Barabási A (2002) Linked: The New Science of Networks. Perseus Books Group.
2. Albert R, Barabási A (2002) Statistical mechanics of complex networks. Rev Mod Phys 74: 47–97.
3. Newman M (2003) The structure and function on complex networks. SIAM Reviews 45: 167.
4. Boccaletti S, Latora V, Moreno Y, Chavez M, Hwang D (2006) Complex networks: Structure and dynamics. Phys Rep 424: 175.
5. Barrat A, Barthelemy M, Vespignani A (2008) Dynamical processes on complex networks. Cambridge: Cambridge University Press.
6. May RM (1972) Will a large complex system be stable? Nature 238: 413.

7. Allesina S, Tang S (2012) Stability criteria for complex ecosystems. Nature 483: 205–208.
8. Atmar W, Paterson BD (1993) The measure of order and disorder in the distribution of species in fragmented habitat. Oecologia 96: 373–382.
9. Wright DH, Reeves JH (1992) On the meaning and measurement of nestedness of species assemblages. OECOLOGIA 92: 414–428.
10. Fischer J, Lindenmayer D (2002) Treating the nestedness temperature calculator as a "black box" can lead to false conclusions. Oikos 99: 193–199.
11. Ulrich W, Almeida-Neto M, Gotelli NJ (2009) A consumer's guide to nestedness analysis. Oikos 118: 3–17.
12. Bascompte J, Jordano P, Melián CJ, Olesen JM (2003) The nested assembly of plant animal mutualistic networks. Proc Nat Acad Sci 100: 9383–9387.
13. Bastolla U, Fortuna M, Pascual-García A, Ferrera A, Luque B, et al. (2009) The architecture of mutualistic networks minimizes competition and increases biodiversity. Nature 458: 1018–21.
14. Sugihara G, Ye H (2009) Cooperative network dynamics. Nature 458: 979.
15. Staniczenko PPA, Kopp J, Allesina S (2013) The ghost of nestedness in ecological networks. Nature Communications 4: 139.
16. Molloy M, Reed B (1995) A critical point for random graphs with a given degree sequence. Random Structures and Algorithms 6: 161–180.
17. Newman M (2002) Mixing patterns in networks. Phys Rev Lett 89: 208701.
18. Dorogovtsev SN, Ferreira AL, Goltsev AV, Mendes JFF (2002010) Zero pearson coefficient for strongly correlated growing treesex network. Physical Review E 81: 031135.
19. Xu X, Zhang J, Sun J, Small M (2009) Revising the simple measures of assortativity in complex networks. Physical Review E 80: 056106.
20. Catanzaro M, Boguñá M, Pastor-Satorras R (2005) Generation of uncorrelated random scale-free networks. Phys Rev E 71: 027103.
21. Maslov S, Sneppen K, Zaliznyak A (2004) Detection of topological patterns in complex networks: Correlation profile of the internet. Physica A 333: 529–540.
22. Park J, Newman M (2003) The origin of degree correlations in the internet and other networks. Phys Rev E 66: 026112.
23. Johnson S, Torres J, Marro J, Muñoz MA (2010) Entropic origin of disassortativity in complex networks. Phys Rev Lett 104: 108702.
24. Wang F, Landau D (2001) Efficient multiple-range random walk algorithm to calculate the density of states. Physical Review Letters 86: 2050–2053.
25. Gotelli N (2001) Research frontiers in null model analysis. Global Ecology and Biogeography 10: 337–343.
26. Burgos E, Ceva H, Perazzo RP, Devoto M, Medan D, et al. (2007) Why nestedness in mutualistic networks? Journal of Theoretical Biology 249: 307–313.
27. Fortuna M, Bascompte J (2006) Habitat loss and the structure of plant-animal mutualistic networks. Ecology Letters 9: 281–286.